\documentstyle[prd,aps,epsfig]{revtex}
\begin{document}
\draft

\title{Superfluid phases of quark matter, II: Phenomenology and sum rules}
\author{Kei Iida$^{1,2,*}$ and Gordon Baym$^{1}$}
\insert\footins{${^*}$Present address: RIKEN, 2-1 Hirosawa, Wako,
Saitama 351-0198, Japan.}
\address{$^{1}$Department of Physics, University of Illinois at
Urbana-Champaign, 1110 West Green Street, Urbana, Illinois 61801-3080}
\address{$^{2}$Department of Physics, University of Tokyo,
7-3-1 Hongo, Bunkyo, Tokyo 113-0033, Japan}
\date{\today}
\maketitle
\begin{abstract}

    We derive sum rules for a uniform, isotropic superfluid quark-gluon plasma
with massless quarks, first laying out the phenomenological equations obeyed
by a color superconductor in terms of macroscopic observables such as the
superfluid mass and baryon densities, and the electric and magnetic gluon
masses, and then expressing these quantities in terms of equilibrium
correlation functions.  From the transverse part of the long wavelength baryon
current-momentum correlation function we derive an exact expression for the
superfluid baryon density, and from the longitudinal part, an $f$-sum
rule.  From the transverse part of the long wavelength {\em color}
current-current correlation function we derive the superfluid Meissner mass,
and from the longitudinal part, the Debye mass.  These masses constrain
integrals of the transverse and longitudinal parts of the gluon propagator
over frequencies, and provide self-consistent conditions for a solution to the
gap equation beyond weak coupling.

\end{abstract}

    Dense degenerate quark matter is expected to exhibit color
superconductivity in the color-antitriplet channel; the predictions are based
on a weak coupling analysis of the gap equation
\cite{barrois,BL,son,hong,SW,PR,dirk,dirk2}, as well as Ginzburg-Landau theory
\cite{BL,I}.  In possible physical realizations in neutron stars and
ultrarelativistic heavy-ion collisions, such a superconductor would be in a
strongly coupled, color-singlet state.  The equilibrium properties of such
superconducting matter have yet to be clarified in detail.  In the strong
coupling regime, Ginzburg-Landau theory delineates the possible phase diagrams
near the critical temperature $T_c$, but quantitative predictions for the
fundamental parameters of the theory are lacking.

    In this paper we derive exact sum rules obeyed by the transverse and
longitudinal momentum-momentum and baryon current-momentum correlation
functions and the gluon propagator, $D$, in color superconductors.  The sum
rules, which are related to the linear response of the equilibrated many-body
system to a current-inducing perturbation, connect the long wavelength
behavior of the correlations with macroscopic observables \cite{pines,baym}.
Such observables include the superfluid mass density, the superfluid baryon
density, and the magnetic mass or inverse penetration depth, in addition to
quantities such as the charge conductivity and the Debye screening length that
play a role in the normal state.  These sum rules act as self-consistency
conditions that must be satisfied by approximate theories of thermodynamics
and correlations in the superconducting state.  To derive the sum rules we
first set up the phenomenological equations obeyed in a relativistic
superfluid plasma in terms of the macroscopic observables, and then turn to
the expressions for these observables in terms of correlation functions.  We
consider a uniform, isotropic color superconductor of three-flavor massless
quarks at finite temperature, $T$, and baryon chemical potential, $\mu_b$, and
use units $\hbar=c=1$.

\section{Phenomenology of relativistic superfluids}

    Let us first review the phenomenological equations obeyed by the baryon
current and momentum density of a relativistic superfluid plasma in
non-dissipative hydrodynamics, linearized about equilibrium with small
velocities.  Through these equations we derive the relativistic relation of
the superfluid mass density, $\rho_s$, and the superfluid baryon density,
$n_s$, quantities familiar in the nonrelativistic context \cite{landau}.
Consistency of the hydrodynamic equations dictates as well the form of the
superfluid acceleration equation in a relativistic superfluid.  The extension
of the relativistic superfluid hydrodynamic equations,
$(\ref{mom})$--$(\ref{ent})$, $(\ref{ene})$, and $(\ref{accel})$ below, to
arbitrary velocities may be found in \cite{khalat}.

    The momentum density, $\bf g$ ($g_i=-T_{0i}$, the off-diagonal components
of the stress tensor), is given in terms of the (small) velocities ${\bf v}_s$
of the superfluid and ${\bf v}_n$ of the normal components by
\begin{equation}
{\bf g} = \rho_s {\bf v}_s+ \rho_n {\bf v}_n\  ,
\label{g}
\end{equation}
where $\rho_s$ is the superfluid mass density, $\rho_n$ is the normal mass
density, and $\rho_s + \rho_n = \rho + P$, where $\rho$ is the total mass
density in the system at rest (the internal energy) and $P$ is the pressure.
That the superfluid velocity is an independent thermodynamic degree of freedom
is a fundamental property of the paired state.  In non-relativistic
superfluids, $\rho + P$ reduces to $mn$, where $m$ is the rest mass of the
carriers, of density $n$; relativistically one must retain the contribution of
the pressure in the momentum density.  Similarly the baryon current is given
in terms of the normal and superfluid velocities by
\begin{equation}
  {\bf j}_b = n_s{\bf v}_s + n_n{\bf v}_n,
  \label{jb}
\end{equation}
where $n_s$ is the superfluid baryon density, $n_n$ the normal baryon density,
and $n_s+n_n=n_b$, the total baryon density.

    The superfluid mass density and superfluid baryon density are closely
related.  As we derive below,
\begin{equation}
   \rho_s = \mu_b n_s,
\label{rhos}
\end{equation}
while the normal density obeys
\begin{equation}
  \rho_n = \mu_b n_n + Ts,
\label{rhon}
\end{equation}
where $s$ is the entropy density.  The latter follows from
Eq.~(\ref{rhos}) together with the relation for the thermodynamic internal
energy density, $\rho = \mu_b n_b + Ts -P$.  Thus
\begin{equation}
{\bf g}= \mu_b {\bf j}_b +Ts{\bf v}_n.
\label{gmu}
\end{equation}

    The basic hydrodynamic equations for the superfluid are the
equations of momentum and baryon conservation, and of entropy flow.  In
linearized hydrodynamics, $\bf g$ is driven by pressure gradients according to
\begin{equation}
\frac{\partial \bf g}{\partial t} + \nabla P = 0.
\label{mom}
\end{equation}
Baryon conservation reads as usual,
\begin{equation}
\frac{\partial n_b}{\partial t} + \nabla\cdot {\bf j}_b  = 0.
\label{bar}
\end{equation}
Since the entropy in a superfluid system is carried only by the normal
fluid, in the absence of dissipation, the entropy density obeys
\begin{equation}
  \frac{\partial s}{\partial t} + \nabla\cdot ({\bf v}_n s) = 0.
\label{ent}
\end{equation}
Furthermore, to second order in the flow velocities, the conserved energy
density, $E = T_{00}$, is given by
\begin{equation}
  E = \frac{\rho_s}{2}v_s^2  + \frac{\rho_n}{2}v_n^2 + \rho,
\label{e}
\end{equation}
and the equation for conservation of energy is
\begin{equation}
\frac{\partial E}{\partial t} + \nabla\cdot {\bf g} = 0.
\label{ene}
\end{equation}

    To derive Eqs.~(\ref{rhos}) and (\ref{rhon}), we explicitly calculate the
time derivative of (\ref{e}), keeping only terms of second order, and use the
above equations, together with the usual first variation, $d\rho = \mu_b dn_b
+ Tds$, and the Gibbs-Duhem relation, $\nabla P = n_b\nabla \mu_b + s \nabla
T$, to find
\begin{equation}
  \frac{\partial E}{\partial t} + \nabla\cdot(\mu_b{\bf j}_b +Ts{\bf v}_n) =
     ({\bf j}_b - n_b {\bf v}_n)\cdot\nabla\mu_b  + \rho_s({\bf v}_s-{\bf v}_n)
      \cdot \frac{\partial {\bf v}_s}{\partial t}.
\label{econs}
\end{equation}
Identifying the energy current with the momentum density, we see that ${\bf g}$
is given by Eq.~(\ref{gmu}), from which Eqs.~(\ref{rhos}) and (\ref{rhon})
follow.

    In addition, the right side of Eq.~(\ref{econs}) must vanish identically.
Eliminating ${\bf j}_b$ there by means of Eq.~(\ref{gmu}) we find as a
necessary condition for this term to vanish that ${\bf v}_s$ obeys the
{\em superfluid acceleration equation},
\begin{equation}
  \mu_b\frac{\partial {\bf v}_s}{\partial t} + \nabla \mu_b  = 0.
  \label{accel}
\end{equation}
The superfluid acceleration equation follows directly from the fact that
the baryon chemical potential and superfluid velocity are given in terms of
the phase $\phi$ of the order parameter by
\begin{equation}
  \frac23 \mu_b = - \frac{\partial \phi}{\partial t},
  \label{muphi}
\end{equation}
and
\begin{equation}
  \frac23 \mu_b {\bf v}_s = \nabla \phi.
   \label{ord}
\end{equation}
The factor 2/3 is the baryon number per pair.  We recall that the order
parameter takes the form $ \Psi_{abfh}(x) \equiv \langle \psi_{af}(x){\bar
\psi}^C_{bh}(x)\rangle=|\Psi_{abfh}(x)|e^{i\phi(x)}$, where $\psi_{af}$ is the
spinor for quarks of color $a$ and flavor $f$, and $\psi^{C}_{af}\equiv C{\bar
\psi}^{T}_{af}$ is the charge-conjugate spinor in the Pauli-Dirac
representation.  In a non-relativistic system, the first $\mu_b$ in
Eq.~(\ref{accel}) and in Eq.~(\ref{ord}) become simply the rest mass, $m$, of
the carriers.

    An important consequence of Eq.~(\ref{ord}) is
that the circulation is quantized according to
\begin{equation}
\oint d{\bf \ell}\cdot \frac23 \mu_b {\bf v}_s = 2\pi \nu,
\end{equation}
where the integral is around any closed path and $\nu$ is an integer.

\section{Momentum and baryon current correlation functions}

    We now derive the phenomenological superfluid densities $\rho_s$ and $n_s$
microscopically in terms of momentum and baryon current correlation functions,
which characterize the linear response of the system to external
disturbances, in particular here, a Galilean transformation.  Consider the
situation, following Ref.~\cite{baym}, in which an infinitely long cylinder
containing a color superconductor moves very slowly with uniform velocity
${\bf v}$ along its axis, which we take to be along $\hat z$.  We assume that
the normal component is in equilibrium with the walls, so that ${\bf v}$
becomes the normal velocity, and that the superfluid component remains at
rest.

    The response of the system to this motion of the walls can be described in
terms of the {\em transverse} response to a static long wavelength
perturbation, $\int d^3{\bf r} {\bf g}({\bf r})\cdot {\bf v}$, where the
momentum density operator is
\begin{equation}
g_i=\sum_{af}\psi_{af}^\dagger(-i\nabla_i\delta_{ab} -\frac g2
\lambda^\alpha_{ab} A^{\alpha i})
      \psi_{bf}+\sum_{\alpha}({\bf E}^\alpha \times{\bf B}^\alpha)_i;
  \label{bc}
\end{equation}
the $A_\mu^\alpha$ are the color gauge fields, with field tensors
$F^\alpha_{\mu\nu}=\partial_\mu A_\nu^\alpha - \partial_\nu A_\mu^\alpha -g
f_{\alpha\beta\gamma} A_\mu^\beta A_\nu^\gamma$ as well as field strengths
$E^\alpha_i = F^{i0}_\alpha$ and $B^\alpha_i = -\frac12 \epsilon_{ijk}
F^{jk}_\alpha$, $g$ is the color coupling constant, and the
$\lambda_{ab}^\alpha$ are the Gell-Mann matrices.

    The induced baryon current is given by
\begin{equation}
  \langle{\bf j}_b \rangle_{\bf v} = \lim_{{\bf k}\to 0} \chi_T^{[jg]}
   ({\bf k},0){\bf v},
   \label{nbgal}
\end{equation}
where the baryon current operator is ${\bf j}_b = \frac13\sum_{af}{\bar
\psi}_{af} \mbox{\boldmath $\gamma$}\psi_{af}$, and $\chi_T^{[jg]}({\bf k},0)$
is the transverse component of the baryon current--momentum density
correlation function \cite{note1},
\begin{equation}
  \chi_{ij}^{[jg]}({\bf k},z)=
   \int_{-\infty}^{\infty} \frac{d\omega}{2\pi}
   \frac{\langle [j_{bi}, g_j] \rangle({\bf k},\omega)}{z-\omega}.
  \label{jg}
\end{equation}
We write here, for general operators $a({\bf r},t)$ and $b({\bf r},t)$,
\begin{equation}
\langle [a,b] \rangle ({\bf k},\omega)
    = -i \int d^3({\bf r}-{\bf r}')\int_{-\infty}^{\infty} d(t-t')
   e^{-i{\bf k\cdot}({\bf r}-{\bf r}')}e^{i\omega(t-t')}
  \langle [a({\bf r},t), b({\bf r}',t')]\rangle,
  \label{bcc1}
\end{equation}
where $\langle\cdots\rangle$ is the ensemble average at given $T$ and
$\mu_b$.  The retarded commutator is given by taking the limit of $z$
approaching the real axis from above in Eq.\ (\ref{jg}).

    Comparing Eqs.~(\ref{nbgal}) and (\ref{jb}) we see then that the normal
baryon density is given in terms of the transverse baryon current--momentum
density correlation function by
\begin{equation}
  n_n = \lim_{{\bf k}\to 0} \chi^{[jg]}_T({\bf k},0).
  \label{nnjg}
\end{equation}
This equation is effectively a sum rule obeyed by $\chi^{[jg]}_T$.  Similarly,
the induced momentum density is given in terms of the transverse momentum
density--momentum density correlation function by
\begin{equation}
  \langle{\bf g} \rangle_{\bf v} = \lim_{{\bf k}\to 0} \chi_T^{[gg]}
   ({\bf k},0){\bf v},
   \label{gggal}
\end{equation}
where for complex frequency, $z$,
\begin{equation}
  \chi_{ij}^{[gg]}({\bf k},z)=
   \int_{-\infty}^{\infty}\frac{d\omega}{2\pi}
   \frac{\langle [g_i, g_j]\rangle(\bf k,\omega)}{z-\omega}.
  \label{bcc}
\end{equation}
Thus, Eq.~(\ref{gggal}) with (\ref{g}) yields the transverse sum rule,
\begin{equation}
  \lim_{{\bf k}\to 0} \chi^{[gg]}_T({\bf k},0)
     = \mu_bn_n + Ts.
  \label{nngg}
\end{equation}

    To derive the longitudinal versions of the sum rules (\ref{nnjg}) and
(\ref{nngg}) we suppose instead that the cylinder is still long but finite
with closed ends.  Then the superfluid component flows together with the
normal component, leading to $\langle {\bf j}_b \rangle_{\bf v} = n_b {\bf v}$
instead of $n_n {\bf v}$.  In this situation, the linear response analysis
yields \cite{baym}
\begin{equation}
  \langle {\bf j}_b\rangle_{\bf v} =
   \lim_{{\bf k}\to 0} \chi_L^{[jg]}({\bf k},0){\bf v}.
   \label{jllr}
\end{equation}
We thus obtain the $f$-sum rule
\begin{equation}
   \chi_{L}^{[jg]}({\bf k},0)  = n_b,
   \label{bls}
\end{equation}
as $|{\bf k}|\to 0$.  Note that the $f$-sum rule (\ref{bls}) can be directly
derived, for general ${\bf k}$, from the baryon conservation law (\ref{bar})
and the equal time commutation relation, $\langle[j_{b0}({\bf r},t), {\bf
g}({\bf r}',t)]\rangle= -i n_b \nabla \delta({\bf r}-{\bf r}')$, where
$j_{b0}= \frac13\sum_{af} \psi_{af}^\dagger \psi_{af}$.
This sum rule can be
rewritten in terms of $\chi^{[gg]}_L$ as
\begin{equation}
   \lim_{{\bf k}\to 0} \chi_{L}^{[gg]}({\bf k},0)  = \mu_bn_b+Ts =
        \rho + P.
   \label{gls}
\end{equation}
To derive this result we calculate ${\bf g}$ from Eq.~(\ref{gmu}) to first
order in ${\bf v}={\bf v}_n = {\bf v}_s$, using $\langle {\bf j}_b
\rangle_{\bf v} = n_b {\bf v}$ and the fact that the entropy term is
explicitly first order in ${\bf v}$.

    The four sum rules (\ref{nnjg}), (\ref{bls}), (\ref{nngg}), and
(\ref{gls}) relate the total and superfluid baryon densities $n_b$ and $n_s$,
and the long wavelength behaviors of the correlation functions $\chi_T^{[jg]}$
$\chi_L^{[jg]}$, $\chi_T^{[gg]}$, and $\chi_L^{[gg]}$.  At $T=0$, as in
ordinary superconductors, longitudinal first sound modes, the only low-lying
excitations for colors and flavors involved in the pairing, contribute only to
$\chi_L^{[jg]}$ and $\chi_L^{[gg]}$, and hence $n_s >0$ \cite{baym}.

\section{Color phenomenology and color current correlation functions}

    Color superconductors have the property of screening out color magnetic
fields, the color Meissner effect.  Following the line of argument of Ref.\
\cite{baym}, we consider the linear response of the system to an applied
static long wavelength transverse color magnetic field, ${\bf A}_{\rm ext}^
\gamma({\bf r})= {\bf A}_{\rm ext}^\gamma({\bf k})e^{i{\bf k \cdot r}}$,
where ${\bf k\cdot}{\bf A}_{\rm ext}^\gamma({\bf k})=0$.  The external
field ${\bf A}_{\rm ext}^\gamma$ produces currents of various colors,
$\beta$, which in turn induce color fields ${\bf A}_{\rm ind}^\beta({\bf r})$;
the total color field is ${\bf A}^\beta = {\bf A}_{\rm ext}^\beta + {\bf
A}_{\rm ind}^\beta$.  In the static long wavelength limit, the induced
transverse color currents are given in terms of the
total transverse color field by the London equation,
\begin{equation}
  \langle {\bf j}^{\alpha T}({\bf r}) \rangle_{{\bf A}_{\rm ext}^\gamma}=
    -(m_M^2)_{\alpha\beta} {\bf A}^\beta ({\bf r}),
\label{london}
\end{equation}
where $m_M$ is the magnetic mass matrix, non-zero in the
paired state.  The inverses of its eigenvalues are the length scales on
which color magnetic fields are screened in the superconductor.

    To linear order in ${\bf A}_{\rm ext}$, the long wavelength induced
currents are given microscopically by
\begin{equation}
  \langle {\bf j}^{\beta T} ({\bf r}) \rangle_{{\bf A}_{\rm ext}^\gamma}=
  -\lim_{{\bf k}\to0}
   \chi_T^{\beta\gamma}({\bf k},0){\bf A}^{\gamma}_{\rm ext}({\bf r}),
  \label{jclr}
\end{equation}
where $\chi_T$ is the transverse part of the color current-current correlation
function,
\begin{equation}
  \chi_{\mu\nu}^{\alpha\beta}({\bf k},z)=
   \int_{-\infty}^{\infty}\frac{d\omega}{2\pi}
   \frac{\langle [j_\mu^\alpha, j_\nu^\beta]\rangle(\bf k,\omega)}{z-\omega},
  \label{ccc}
\end{equation}
and the color current operator for gluonic index $\alpha$ is
\begin{equation}
  j_\mu^\alpha=\frac12 g \sum_{abf}{\bar\psi}_{af} \lambda_{ab}^\alpha
               \gamma_\mu \psi_{bf}
               -gf_{\alpha\beta\gamma}A^{\beta\nu} F^\gamma_{\mu\nu}.
  \label{cc}
\end{equation}
The linearized field equation for ${\bf A}_{\rm ind}^\beta$,
\begin{equation}
  \nabla\times(\nabla\times{\bf A}_{\rm ind}^\beta) =
  \langle {\bf j}^{\beta T}({\bf r})\rangle_{{\bf A}_{\rm ext}^\gamma},
   \label{max}
\end{equation}
implies that the total color field is
\begin{equation}
  {\bf A}^\beta({\bf k}) = [(\varepsilon^T)^{-1}]_{\beta\gamma}
  {\bf A}_{\rm ext}^\gamma({\bf k}),
    \label{abeta}
\end{equation}
where $\varepsilon^T_{\alpha\beta}({\bf k})$ is the static transverse color
dielectric function, defined by
\begin{equation}
  [(\varepsilon^T)^{-1}]_{\alpha\beta}({\bf k})
     = \delta_{\alpha\beta} -
         {\chi}_T^{\alpha\beta}({\bf k},0)/|{\bf k}|^2.
\end{equation}
One can write $\varepsilon^T_{\alpha\beta}({\bf k})$ in terms of the
screened correlation function, ${\tilde\chi}_T^{\alpha\beta}({\bf k},0) \equiv
\chi_T^{\alpha\gamma}({\bf k},0)\varepsilon^T_{\gamma\beta}({\bf k})$ (the
irreducible bubble), as
\begin{equation}
  \varepsilon^T_{\alpha\beta}({\bf k})
     = \delta_{\alpha\beta} +
        {\tilde\chi}_T^{\alpha\beta}({\bf k},0)/|{\bf k}|^2.
\label{screen}
\end{equation}
The magnetic mass matrix is thus given in terms of $\tilde\chi_T$ by
\begin{equation}
  (m_M^2)_{\alpha\beta} = \lim_{{\bf k}\to0}
   {\tilde\chi}_T^{\alpha\beta} ({\bf k},0).
\end{equation}

    The transverse screening lengths are closely related to the superfluid
baryon density $n_s$.  We may see the explicit relation near $T_c$, where we
derived the equilibrium properties utilizing general Ginzburg-Landau theory
\cite{I,II}.  Within color and flavor antisymmetric pairing channels having
zero total angular momentum, even parity, and aligned chirality, which include
the two-flavor and color-flavor locked condensates as optimal states, we
obtain the London equation  for the induced current densities
(\ref{jclr}) as (see Appendix and \cite{II})
\begin{equation}
  \langle {\bf j}^{\alpha T} ({\bf r})\rangle_{{\bf A}_{\rm ext}^\gamma}
= - K_T \left(\frac g2\right)^2 {\rm Re}\left\{{\rm Tr}\left[
   \left((\lambda^\alpha)^* \phi_+ + \phi_+ \lambda^\alpha\right)
   \left((\lambda^\beta)^* \phi_+ + \phi_+ \lambda^\beta\right)^\dagger
                 \right]_F \right\}  {\bf A}^\beta({\bf r})
   \label{jcgl}
\end{equation}
with coefficient
\begin{equation}
   K_T=\frac{9n_s}{4\mu_b {\rm Tr}(\phi_+^\dagger \phi_+)_F}.
   \label{k}
\end{equation}
Here $(\phi_+)_{abfh}$ is the pairing gap of the quark of color $a$ and
flavor $f$ with that of color $b$ and flavor $h$, and the subscript $F$
denotes the gap calculated for the paired quarks lying on the Fermi surfaces;
for a color neutral system, the Fermi energies reduce to a single value
$\mu_b/3$ as $T$ approaches $T_c$.  The relation (\ref{k}) is basically that
obtained by Josephson \cite{josephson}, $\rho_s =A_{\bot}(m/\hbar)^2|\Psi|^2$,
for superfluid He II.  Equations (\ref{london}) and (\ref{jcgl}) then yield
the relation between the magnetic mass, the superfluid baryon density, and the
order parameter near $T_c$:
\begin{equation}
  (m_M^2)_{\alpha\beta}
  = \frac{9 g^2 n_s}{16\mu_b }
    \frac{{\rm Re}{\rm Tr}\left[
   \left((\lambda^\alpha)^*\phi_+ + \phi_+ \lambda^\alpha\right)
   \left((\lambda^\beta)^*\phi_+ + \phi_+ \lambda^\beta \right)
       ^\dagger \right]_F}{{\rm Tr}(\phi_+^\dagger \phi_+)_F}.
    \label{meissner}
\end{equation}

    For the color-flavor locked phase, where $(\phi_+)_{abfh}=
\kappa_A(\delta_{af}\delta_{bh}-\delta_{ah}\delta_{bf})$, we obtain
\begin{equation}
  K_T = \frac{3 n_s}{16\mu_b|\kappa_A|^2_F},
\end{equation}
and
\begin{equation}
  (m_M^2)_{\alpha\alpha} = \frac{3g^2 n_s}{8\mu_b}.
\end{equation}
Similarly, for the two-flavor channel, where $(\phi_+)_{abfh}=\epsilon_{fhs}
\epsilon_{abc}d_c$ ($s$, the strange flavor),
\begin{equation}
  K_T = \frac{9 n_s}{16\mu_b|{\bf d}|^2_F},
\end{equation}
and
\begin{equation}
 (m_M^2)_{\alpha\alpha} = \left\{
 \begin{array}{ll}
  0, & \quad \mbox{$\alpha=1,2,3,$} \\
  9g^2 n_s/16\mu_b, & \quad \mbox{$\alpha=4,5,6,7,$} \\
  3g^2 n_s/4\mu_b, & \quad \mbox{$\alpha=8.$}
 \end{array}
\right.
\end{equation}

    In contrast to the behavior of transverse color fields, color longitudinal
fields are screened in both the normal and superconducting states.
Longitudinal color charge correlations act to expel low frequency longitudinal
color fields as in the response of nonrelativistic electron systems to an
external longitudinal electromagnetic field \cite{pines}.  We can see this
behavior by simply replacing the transverse external field ${\bf
A}^{\gamma}_{\rm ext}$ above with a slowly varying
longitudinal one, satisfying ${\bf k}\times {\bf A}^{\gamma}_{\rm ext}({\bf
k})=0$, with time dependence $e^{-i(\omega+i\eta)t}$ (here, $\omega \simeq 0$
and $\eta$ is a positive infinitesimal).  Then, as in the derivation of Eq.\
(\ref{abeta}), we obtain, for the total color longitudinal fields in a gauge
where the scalar fields $A_0^\gamma$ vanish,
\begin{equation}
  {\bf A}^\beta({\bf k},\omega) = [(\varepsilon^L)^{-1}]_{\beta\gamma}
  {\bf A}_{\rm ext}^\gamma({\bf k},\omega),
    \label{abetal}
\end{equation}
where $\varepsilon^L_{\alpha\beta}({\bf k},\omega+i\eta)\equiv
\delta_{\alpha\beta}-{\tilde\chi}_L^{\alpha\beta}({\bf k},\omega+i\eta) /|{\bf
k}|^2$, with the screened correlation function ${\tilde\chi}_L^{\alpha
\beta}({\bf k},\omega+i\eta)=\chi_L^{\alpha\gamma}({\bf k},\omega+i\eta)
\varepsilon^L_{\gamma\beta}({\bf k},\omega+i\eta)$, is the longitudinal color
dielectric function.  [Here it is more convenient to write
$\tilde\chi_L^{\alpha\beta} \equiv \tilde\chi_{00}^{\alpha\beta}$, in
constrast to the definition of the longitudinal part in note \cite{note1}.]

    In the static long wavelength limit, the longitudinal correlation function
reduces to minus the square of the electric mass tensor,
\begin{equation}
  \lim_{{\bf k}\to0}
  {\tilde\chi}_L^{\alpha\beta}({\bf k},0)  = - (m_E^2)_{\alpha\beta};
\end{equation}
then ${\bf A}^\alpha({\bf k},0) \to |{\bf k}|^2 (m_E^{-2})_{\alpha\beta}
{\bf A}_{\rm ext}^\beta({\bf k},0),$ so that the total longitudinal field is
screened in the long wavelength limit.  In the opposite limit of spatially
uniformity, with slow variation in time,
\begin{equation}
  \lim_{\omega\to0}\lim_{{\bf k}\to0} (\omega/|{\bf k}|)^2
   {\tilde\chi}_L^{\alpha\beta}({\bf k},\omega+i\eta) =
    (\omega_p^2)_{\alpha\beta},
\end{equation}
where $\omega_p^{\alpha\beta}$ is the plasma frequency matrix; in this limit
${\bf A}^\alpha(0,\omega) \to -\omega^2 (\omega_p^{-2})_{\alpha\beta}
{\bf A}_{\rm ext}^\beta(0,\omega)$.

    Near $T_c$, the square of the plasma frequency is given by the sum of
normal and superconducting contributions, $(\omega_p^2)_{\alpha\beta} =
\delta_{\alpha\beta}\omega_p^2 + (m_M^2)_{\alpha\beta}$ (see Appendix); in
weak coupling, $\omega_p^2 = g^2 \mu_b^2/18\pi^2 + g^2 T^2 /2$ \cite{chin} and
$K_T = 7\zeta(3)n_b/16\pi^2 T_c^2 \mu_b$ \cite{BL,IH}, where $\zeta(3)=1.202 
\ldots$ is the Riemann zeta function.  The deviation of 
$\omega_p^{\alpha\beta}$ from $\delta_{\alpha\beta}\omega_p$ below $T_c$ comes
from the fact, as clarified by Rischke $et$ $al.$ \cite{dirk,RSS} for the 
two-flavor channel, that the condensate changes the color dielectric properties
in such a way that the color superconductor is transparent to the low energy 
gluons having color charge associated with the two colors carried by a Cooper 
pair.  In weak coupling, $m_E^{\alpha\beta}$ reduces above $T_c$ to the usual 
Debye mass $\sqrt3\omega_p\delta_{\alpha\beta}$ \cite{chin}.  Below $T_c$, 
however, $m_E^{\alpha\beta}$ generally deviates from the normal Debye mass, 
due to the modification of the color dielectric properties by the condensate
\cite{dirk,RSS}.  Just below $T_c$, $(m_E^2)_{\alpha\beta}$ behaves as
$3\omega_p^2 \delta_{\alpha\beta} -3(m_M^2)_{\alpha\beta}$ (see Appendix).

    The relations of the Meissner and Debye screening masses, $m_M$ and $m_E$,
to the color current--current correlation functions can be cast in terms of
sum rules obeyed by the gluon propagator,
\begin{equation}
   D_{\mu\nu}^{\alpha\beta}({\bf r}t,{\bf r'}t') =
     -i\langle T[A_\mu^\alpha({\bf r},t)A_\nu^\beta({\bf r'},t')]\rangle.
\end{equation}
Since this propagator dominates the infrared structure of the gap equation
\cite{PR}, the sum rule constraints on the propagator are important to take
into account in constructing a self-consistent solution to the gap equation
through inclusion of polarization effects of the superconducting medium.

    The transverse propagator $D_T$, with spectral representation,
\begin{equation}
   D_T^{\alpha\beta}({\bf k},z) =
   \int_{-\infty}^{\infty} \frac{d\omega}{2\pi}
   \frac{B_T^{\alpha\beta}({\bf k},\omega)}{z-\omega},
  \label{DTB}
\end{equation}
is related to the transverse part, ${\tilde\chi}_T^{\alpha\beta}({\bf
k},z)$, of the irreducible current-current correlation function by
\begin{equation}
  (D_T^{-1})_{\alpha\beta}({\bf k},z) \equiv \delta_{\alpha\beta}(z^2-|{\bf
     k}|^2)- {\tilde\chi}_T^{\alpha\beta}({\bf k},z).
  \label{dysont}
\end{equation}
Thus taking $z=0$ and the limit of small $|{\bf k}|$, we derive the
transverse sum rule,
\begin{equation}
   \lim_{{\bf k}\to0}\int_{-\infty}^{\infty} \frac{d\omega}{2\pi}
   \frac{B_T^{\alpha\beta}({\bf k},\omega)}{\omega} =
       (m_M^{-2})_{\alpha\beta}.
  \label{dtsum}
\end{equation}

    Similarly, the longitudinal propagator, $D_L=D_{00}$ in the radiation
gauge, has the spectral representation,
\begin{equation}
   D_L^{\alpha\beta}({\bf k},z) =   \frac{1}{|{\bf k}|^2}\delta_{\alpha\beta}
     + \int_{-\infty}^{\infty} \frac{d\omega}{2\pi}
   \frac{B_L^{\alpha\beta}({\bf k},\omega)}{z-\omega},
  \label{DLB}
\end{equation}
and is given in terms of the longitudinal part of the irreducible
current-current correlation function by
\begin{equation}
  (D_L^{-1})_{\alpha\beta}({\bf k},z)
    =\delta_{\alpha\beta}|{\bf k}|^2-{\tilde\chi}_L^{\alpha\beta}({\bf k},z).
  \label{dysonl}
\end{equation}
Again in the static long wavelength limit, we find the longitudinal sum rule:
\begin{equation}
  \lim_{{\bf k}\to0}\left(
       \frac{1}{|{\bf k}|^2}\delta_{\alpha\beta}
      -\int_{-\infty}^{\infty} \frac{d\omega}{2\pi}
   \frac{B_L^{\alpha\beta}({\bf k},\omega)}{\omega}\right)
    = (m_E^{-2})_{\alpha\beta}.
  \label{dlsum}
\end{equation}

    In the normal state the sum rules (\ref{dtsum}) and (\ref{dlsum})
reproduce the results analyzed by Pisarski and Rischke \cite{PR} in terms of
the spectral representation up to one-loop order.  The Meissner masses vanish
in the transverse sector; Landau diamagnetism leads only to $|{\bf k}|^2$
corrections, and not a nonzero screening mass in the limit ${\bf k}\to0$.
The right side of Eq.~(\ref{dtsum}) is replaced by $\delta_{\alpha\beta}
\lim_{{\bf k}\to 0}|{\bf k}|^{-2}$ to leading order in $g$, as in Ref.
\cite{PR}.  The longitudinal sector contains the usual Debye screening, which
is characterized by $(m_E^2)_{\alpha\beta} = 3\omega_p^2\delta_{\alpha\beta}$
to leading order in $g$.

    Rischke \cite{dirk} explicitly calculated the zero temperature, low
frequency, long wavelength limit of the color current-current correlation
function for the two-flavor and color-flavor locked condensates to leading
order in $g$.  Repeating his calculations for three flavors, we find the
screening masses (with no sum over $\alpha$),
\begin{equation}
  (m_E^2)_{\alpha\alpha}=3(m_M^2)_{\alpha\alpha}=
   \frac{21-8\ln 2}{18}\omega_p^2 ,~~ \alpha=1,\ldots,8
   \label{cfl}
\end{equation}
for color-flavor locking, and
\begin{equation}
  \begin{array}{lll}
  (m_E^2)_{\alpha\alpha}=\omega_p^2, & (m_M^2)_{\alpha\alpha} = 0, &
 \alpha =1,2,3 \\
  (m_E^2)_{\alpha\alpha}=2\omega_p^2,& (m_M^2)_{\alpha\alpha}=\omega_p^2/3, &
 \alpha=4,5,6,7 \\
  (m_E^2)_{88}= 3\omega_p^2, & (m_M^2)_{88} = 2\omega_p^2/9 &
  \end{array}
\label{is}
\end{equation}
for the two-flavor channel in which only quarks of color $R$ and $G$, which 
couple to gluons of $\alpha=1,2,3$, undergo BCS pairing.  (The choice of the 
two colors involved in the pairing is arbitrary under the constraint of 
overall color neutrality.)  The screening masses depend on color charge only 
in the two flavor condensate, because there the order parameter is 
{\em anisotropic} in color space in contrast to that in the color-flavor 
locked condensate.  How static transverse screening influences the pairing gap
compared with Landau damping of color magnetic gluons \cite{BMPR} depends 
sensitively on the gluon energies and momenta controlling the pairing gap 
\cite{dirk}.

    Weak coupling calculations ignoring the effects of the superconducting
medium yield the logarithm of the gap \cite{hong,SW,PR},
\begin{equation}
\ln(\Delta/\mu_b)= - 3\pi^2 /\sqrt2 g- 5\ln g + \ldots.
\label{wcs}
\end{equation}
The weak coupling solution is not self-consistent in the sense that it 
satisfies the sum rules (\ref{dtsum}) and (\ref{dlsum}) with the masses 
$m_E^{\alpha\beta}=\sqrt3\omega_p\delta_{\alpha\beta}$ and 
$m_M^{\alpha\beta}=0$, corresponding to the thermodynamics of the weakly 
interacting normal gas \cite{PR}.  As Rischke \cite{dirk2} showed to one loop 
order, the superconducting medium significantly modifies the gluon self-energy
from the normal medium value only in the energy range $|\omega| \lesssim 
\Delta$, which is not sufficient to change the logarithm of the gap from Eq.\ 
(\ref{wcs}) up to subleading order in $g$.  On the other hand, a 
self-consistent solution that satisfies the sum rules (\ref{dtsum}) and 
(\ref{dlsum}) with masses given by Eqs.\ (\ref{cfl}) and (\ref{is}) would 
include contributions in all orders.  The sum rules thus provide a check on 
approximate theories for the pairing gap beyond weak coupling.

    In summary, we have derived the transverse and longitudinal sum rules for
a color superconductor that is uniform and isotropic in ordinary space.  In
doing so we have brought out the relation of the long wavelength behaviors of
the momentum-momentum, baryon current-momentum, and color current correlation
functions to macroscopic quantities such as the superfluid density and the
Meissner and Debye masses, as well as the relevance of the sum rules for the
gluon propagator to the self-consistent solution to the gap equation.  The sum
rules hold for any number of flavors and quark masses.  It is straightforward
to extend the present analysis for color current-current correlations to the
case in which the electric currents coexist with the color currents; in this
situation the electric currents modify the supercurrents through a mixing of
color and electric charge (see, e.g., Refs.\ \cite{II,gorbar}).

     We thank Dirk Rischke for helpful comments.  This work was supported in 
part by a Grant-in-Aid for Scientific Research provided by the Ministry of 
Education, Science, and Culture of Japan through Grant No.\ 10-03687, and in 
part by National Science Foundation Grant No.\ PHY98-00978 and PHY00-98353.

\section*{Appendix: The Ginzburg-Landau region}

    In this Appendix, we summarize the derivation of the London equation
(\ref{jcgl}) near $T_c$, for color and flavor antisymmetric pairing channels
having zero total angular momentum, even parity, and aligned chirality; full
details will be given in Ref.\ \cite{II}.  In the presence of applied weak
color fields, $A_{\rm ext}^{\alpha\mu}(x)$, of very long wavelength and low
frequency, the resultant gradient of the order parameter adds a small
correction to the homogeneous part of the Ginzburg-Landau free energy derived
in Ref.\ \cite{I}.  Up to second order in the gap, this energy correction is
\begin{equation}
  \Omega_{g}= \frac12 K_T {\rm Tr}[(D_i \phi_+)^{\dagger}D_i \phi_{+}]_F
  +\frac12 K_L {\rm Tr}[(D_0 \phi_+)^{\dagger}D_0 \phi_{+}]_F,
   \label{omegag1}
\end{equation}
where the covariant derivative is $D_\mu \phi_+ \equiv \partial_\mu \phi_+
- \frac 12 ig[(\lambda^\alpha)^* \phi_+ + \phi_+ \lambda^\alpha] A_\mu^\alpha$
with the total color fields $A_\mu^\alpha$.  The time dependence of the gap
is, by definition, measured with respect to that in the equilibrium phase,
i.e., with respect to the phase factor $e^{-2i\mu_b t/3}$ arising in the order
parameter from Eq.\ (\ref{muphi}) (see Ref.\ \cite{I}).  As we show here, the
coefficient $K_T$ is given by Eq.~(\ref{k}).  The coefficient $K_L$ is not
necessarily equal to $K_T$ since Lorentz invariance is broken in a many
particle system.  In weak coupling, $K_L$ reduces to $3K_T$ \cite{dirk}.  As
required, $\Omega_{g}$ is invariant under global $U(1)$ gauge transformations
and flavor rotations, as well as under local color $SU(3)$ gauge
transformations.

    To derive the dependence of the coefficient $K_T$ on the superfluid baryon
density $n_s$, Eq.\ (\ref{k}), we consider the situation in the absence of
color fields, in which the pairs move uniformly with small constant velocity
${\bf v}_s$ and the normal fluid remains at rest.  The phase factor of the gap
in the fixed frame transforms by $\phi_+ \to e^{i{\bf P\cdot r}}\phi_+$, where
${\bf P}$ is the total pair momentum.  The total momentum of the superfluid is
then
\begin{equation}
 3n_s {\bf P}/2 =  \rho_s{\bf v}_s.
\label{pvs}
\end{equation}
In this situation,
\begin{equation}
  \Omega_{g}= \frac12 K_T {\bf P}^2 {\rm Tr}(\phi_{+}\phi_+^{\dagger})_F.
   \label{omegag2}
\end{equation}
Following de Gennes \cite{dG2} and W{\" o}lfle \cite{wolfle}, we then
obtain the baryon current density from the usual canonical equation for the
gradient energy density (\ref{omegag2}) as
\begin{eqnarray}
{\bf j}_s = \frac 23 \frac{\delta \Omega_g}{\delta {\bf P}}
= \frac 23 K_T {\rm Tr}(\phi_+^\dagger \phi_+)_F{\bf P}.
\label{js}
\end{eqnarray}
Using Eq.~(\ref{pvs}) to eliminate ${\bf P}$, and Eqs.\ (\ref{jb}) and
(\ref{rhos}), we derive Eq.\ (\ref{k}).  Note that the extra time variation of
the gap leads to terms in  $\Omega_{g}$ of order ${\bf P}^4$, which play no 
role here.

    The color current densities induced by the applied weak color fields near
$T_c$ are composed of the superfluid and normal contributions,
$j^{\alpha\mu}_s$ and $j^{\alpha\mu}_n$.  The superfluid color currents can be
calculated as
\begin{eqnarray}
j_s^{\alpha\mu} &=& \frac{\delta \Omega_g}{\delta A_\mu^\alpha}
\nonumber \\
&=& -[K_T+\delta_{\mu0}(K_L-K_T)]
     \frac g2 {\rm Im}\left\{{\rm Tr}\left[\left((\lambda^\alpha)^*
                          \phi_+ + \phi_+ \lambda^\alpha \right)^\dagger
                          \partial_\mu \phi_+\right]_F \right\}
\nonumber \\
& & + [K_T+\delta_{\mu0}(K_L-K_T)]
   \left(\frac g2\right)^2 A^\beta_\mu {\rm Re}\left\{{\rm Tr}\left[
   \left((\lambda^\alpha)^* \phi_+ + \phi_+ \lambda^\alpha\right)
   \left((\lambda^\beta)^* \phi_+ + \phi_+ \lambda^\beta\right)^\dagger
                 \right]_F \right\}.
\label{cmax}
\end{eqnarray}
Equation (\ref{cmax}) reduces to the London equation (\ref{jcgl}) for
applied static transverse color fields when the spatial variation is of
sufficiently long wavelength that we can ignore the term containing
$\partial_\mu \phi_+$.

    We conclude this Appendix by considering the linear response to applied
longitudinal color fields near $T_c$.  For static color fields of very long
wavelength the induced color density $j_s^{\alpha0}$ can be obtained from Eq.\
(\ref{cmax}) as $j_s^{\alpha0}= (K_L/K_T)(m_M^2)_{\alpha\beta}A^{\beta0}$
where $m_M^2$ is given by Eq.\ (\ref{meissner}).  In weak coupling, this color
density, together with the normal contribution $-3\omega_p^2 A^{\alpha0}$,
leads to the square of the Debye screening mass matrix,
$(m_E^2)_{\alpha\beta}=3\omega_p^2 \delta_{\alpha\beta}
-3(m_M^2)_{\alpha\beta}$.  For uniform color fields varying very slowly in
time, we calculate the induced supercurrent from Eq.\ (\ref{cmax}) as ${\bf
j}_s^{\alpha}=-(m_M^2)_{\alpha\beta} {\bf A}^\beta$, where $m_M^2$ is again
given by Eq.\ (\ref{meissner}).  Combining this supercurrent with the induced
normal current $-\omega_p^2{\bf A}^\alpha$, we obtain the square of the plasma
frequency matrix, $(\omega_p^2)_{\alpha\beta} = \delta_{\alpha\beta}\omega_p^2
+ (m_M^2)_{\alpha\beta}$.

\end{document}